\documentstyle[epsfig,aps]{revtex}
\font\eightit=cmti8                     
\def\r#1{\ignorespaces $^{#1}$}         
\newcommand{\PT}{{\rm P}_{\!\!\scriptscriptstyle\rm T}}
\newcommand{\ET}{{\rm E}_{\scriptscriptstyle\rm T}}
\newcommand{\MET}{\mbox{$\raisebox{.3ex}{$\not$}\ET$}}
\newcommand{\qqbar}{q\bar{q}}
\newcommand{\ppbar}{p\bar{p}}
\newcommand{\ttbar}{t\bar{t}}
\newcommand{\bbar}{b\bar{b}}
\newcommand{\Mtt}{M_{\ttbar}}
\newcommand{\zpr}{Z^{\prime}}
\newcommand{\Mz}{M_{\zpr}}
\newcommand{\bbbar}{b\bar{b}}

\newcommand{\ipb}{ {\rm pb}^{-1} }
\def\Mx{M_{\tiny{\mbox X}}}
\def \gev {{\rm GeV/}c{\rm ^2}}
\def \tev {{\rm TeV/}c{\rm ^2}}
\def\sbr{\sigma_{\tiny {\mbox X}} \cdot \mbox{BR}\{{\mbox X} \rightarrow \ttbar\}}

\def\r#1{\ignorespaces $^{#1}$}
\font\eightit=cmti8
\begin{document}
\draft
\onecolumn
\title{
\begin{center}
Search for New Particles Decaying to $\ttbar$ in $\ppbar$
Collisions at $\sqrt{s}=1.8$~TeV
\end{center}
      }
%
\author{
\hfilneg
\begin{sloppypar}
\noindent
T.~Affolder,\r {21} H.~Akimoto,\r {43}
A.~Akopian,\r {36} M.~G.~Albrow,\r {10} P.~Amaral,\r 7 S.~R.~Amendolia,\r {32} 
D.~Amidei,\r {24} K.~Anikeev,\r {22} J.~Antos,\r 1 
G.~Apollinari,\r {36}
% ...
\end{sloppypar}
\vskip .026in
\begin{center}
(CDF Collaboration)
\end{center}
}

\vskip .026in
\address{
\begin{center}
\r 1  {\eightit Institute of Physics, Academia Sinica, Taipei, Taiwan 11529, 
Republic of China} \\
\r 2  {\eightit Argonne National Laboratory, Argonne, Illinois 60439} \\
\r 3  {\eightit Istituto Nazionale di Fisica Nucleare, University of Bologna,
I-40127 Bologna, Italy} \\
\r 4  {\eightit Brandeis University, Waltham, Massachusetts 02254} \\
\r 5  {\eightit University of California at Los Angeles, Los 
Angeles, California  90024} \\  
\r 6  {\eightit Instituto de Fisica de Cantabria, University of Cantabria, 
39005 Santander, Spain} \\
\r 7  {\eightit Enrico Fermi Institute, University of Chicago, Chicago, 
Illinois 60637} \\
\r 8  {\eightit Joint Institute for Nuclear Research, RU-141980 Dubna, Russia}
\\
\r 9  {\eightit Duke University, Durham, North Carolina  27708} \\
\r {10}  {\eightit Fermi National Accelerator Laboratory, Batavia, Illinois 
60510} \\
\r {11} {\eightit University of Florida, Gainesville, Florida  32611} \\
\r {12} {\eightit Laboratori Nazionali di Frascati, Istituto Nazionale di Fisica
               Nucleare, I-00044 Frascati, Italy} \\
\r {13} {\eightit University of Geneva, CH-1211 Geneva 4, Switzerland} \\
\r {14} {\eightit Harvard University, Cambridge, Massachusetts 02138} \\
\r {15} {\eightit Hiroshima University, Higashi-Hiroshima 724, Japan} \\
\r {16} {\eightit University of Illinois, Urbana, Illinois 61801} \\
\r {17} {\eightit The Johns Hopkins University, Baltimore, Maryland 21218} \\
\r {18} {\eightit Institut f\"{u}r Experimentelle Kernphysik, 
Universit\"{a}t Karlsruhe, 76128 Karlsruhe, Germany} \\
\r {19} {\eightit Korean Hadron Collider Laboratory: Kyungpook National
University, Taegu 702-701; Seoul National University, Seoul 151-742; and
SungKyunKwan University, Suwon 440-746; Korea} \\
\r {20} {\eightit High Energy Accelerator Research Organization (KEK), Tsukuba, 
Ibaraki 305, Japan} \\
\r {21} {\eightit Ernest Orlando Lawrence Berkeley National Laboratory, 
Berkeley, California 94720} \\
\r {22} {\eightit Massachusetts Institute of Technology, Cambridge,
Massachusetts  02139} \\   
\r {23} {\eightit Institute of Particle Physics: McGill University, Montreal 
H3A 2T8; and University of Toronto, Toronto M5S 1A7; Canada} \\
\r {24} {\eightit University of Michigan, Ann Arbor, Michigan 48109} \\
\r {25} {\eightit Michigan State University, East Lansing, Michigan  48824} \\
\r {26} {\eightit University of New Mexico, Albuquerque, New Mexico 87131} \\
\r {27} {\eightit The Ohio State University, Columbus, Ohio  43210} \\
\r {28} {\eightit Osaka City University, Osaka 588, Japan} \\
\r {29} {\eightit University of Oxford, Oxford OX1 3RH, United Kingdom} \\
\r {30} {\eightit Universita di Padova, Istituto Nazionale di Fisica 
          Nucleare, Sezione di Padova, I-35131 Padova, Italy} \\
\r {31} {\eightit University of Pennsylvania, Philadelphia, 
        Pennsylvania 19104} \\   
\r {32} {\eightit Istituto Nazionale di Fisica Nucleare, University and Scuola
               Normale Superiore of Pisa, I-56100 Pisa, Italy} \\
\r {33} {\eightit University of Pittsburgh, Pittsburgh, Pennsylvania 15260} \\
\r {34} {\eightit Purdue University, West Lafayette, Indiana 47907} \\
\r {35} {\eightit University of Rochester, Rochester, New York 14627} \\
\r {36} {\eightit Rockefeller University, New York, New York 10021} \\
\r {37} {\eightit Rutgers University, Piscataway, New Jersey 08855} \\
\r {38} {\eightit Texas A\&M University, College Station, Texas 77843} \\
\r {39} {\eightit Texas Tech University, Lubbock, Texas 79409} \\
\r {40} {\eightit Istituto Nazionale di Fisica Nucleare, University of Trieste/
Udine, Italy} \\
\r {41} {\eightit University of Tsukuba, Tsukuba, Ibaraki 305, Japan} \\
\r {42} {\eightit Tufts University, Medford, Massachusetts 02155} \\
\r {43} {\eightit Waseda University, Tokyo 169, Japan} \\
\r {44} {\eightit University of Wisconsin, Madison, Wisconsin 53706} \\
\r {45} {\eightit Yale University, New Haven, Connecticut 06520} \\
\end{center}
}
\maketitle
\begin{abstract}
We use 106 $\ipb$ of data collected with the Collider Detector at
Fermilab to search for narrow-width, vector particles decaying to a
top and an anti-top quark.  
Model independent upper
limits on the cross section for narrow, vector resonances decaying to $\ttbar$ 
are presented.  At the 95\% confidence level, we exclude the existence of a
leptophobic $\zpr$ boson
in a model
of topcolor-assisted technicolor with
mass $M_{\zpr}$ $<$ 480 $\gev$ for natural width $\Gamma$ = 0.012 $M_{\zpr}$,
and $M_{\zpr}$ $<$ 780 $\gev$ for $\Gamma$ = 0.04 $M_{\zpr}$.
\end{abstract}
\pacs{PACS numbers: 14.65.Ha, 13.85.Ni, 13.85.Qk}
\twocolumn
%
%
\par
In this letter, we present a model-independent search for narrow, vector 
resonances decaying to $\ttbar$.  
This search is sensitive to, for example, a $\zpr$
predicted by topcolor-assisted technicolor\cite{topcolor,models}.  This model
anticipates that the explanation of spontaneous electroweak symmetry breaking
is related to the observed fermion masses, and that the large value of the
top quark mass suggests the introduction of new strong dynamics into the
standard model.  It accounts for the large top quark mass by predicting the
existence of a residual global symmetry SU(3)$\times$U(1) at energies below 1
TeV.  
The SU(3) results in the generation of topgluons which we have searched 
for previously in the $\bbbar$ channel\cite{topgluons}. The U(1) gives the 
$\zpr$ we search for here.  
In one model\cite{models}, the $\zpr$ decays exclusively to quarks 
(leptophobic) resulting in a large cross section for $\ttbar$.

With the $z$-axis defined along the proton beam, 
the Collider Detector at Fermilab (CDF) coordinate system defines 
$\phi$ as the azimuthal angle in the 
transverse plane, $\theta$ as the polar angle, and pseudorapidity $\eta$ as 
$-\ln{(\tan{\frac{\theta}{2}})}$.  Tracking chambers, immersed in a 1.4-Tesla 
solenoidal magnetic field, are used for the detection of charged
particles and the measurement of their momenta.  The precision track
reconstruction of the silicon microstrip
vertex detector (SVX), located immediately outside the beampipe, is used for the
detection of displaced secondary vertices resulting from $b$-quark decays.  
Outside the SVX is the vertex time projection chamber (VTX) which provides 
further tracking information for $|\eta|$ $\leq$ $3.25$.  Both
the SVX and VTX are housed within the central tracking chamber (CTC), a wire
drift chamber used to measure charged particle momenta.  Electromagnetic and
hadronic calorimeters, located beyond the CTC and superconducting solenoid,
measure energy in segmented $\eta$-$\phi$ towers out to $|\eta|$ $<$ 4.2.  Drift
chambers used for muon detection reside outside the calorimetry.  A more 
detailed description of the CDF detector can be found 
elsewhere\cite{detector,prd94}.  

Standard model $\ttbar$ production in
$\ppbar$ collisions at a center of mass energy of $\sqrt{s}=1.8$~TeV is
dominated by $\qqbar$ annihilation, while $\sim$10\% is attributable to
gluon-gluon fusion.  Once a $\ttbar$ pair is produced, each of the top quarks is 
expected to decay almost exclusively to $Wb$.  
The search presented here focuses on the $\ttbar$
event topology in which one $W$ boson decays hadronically while the other
decays to an electron or muon and its corresponding neutrino.  The 
fragmentation of the 
$b$-quarks, as well as the hadronic daughters of the $W$ boson, form 
jets.  Accordingly, $\ttbar$ candidates in this \lq\lq 
lepton + jets" channel are characterized by a single lepton, missing 
transverse energy, $\MET$ \cite{met_def}, due to 
the undetected neutrino, and at least four jets.   Furthermore, a jet resulting 
from a $b$-quark can be identified (or \lq\lq tagged") as such by the 
reconstruction of a secondary vertex from the $b$ hadron decay using the SVX, 
or by using the soft lepton tagging (SLT) algorithm to find an additional 
lepton from a semileptonic $b$ decay\cite{prd94,t_discov}.

Like the top quark mass measurement\cite{top_prl}, events included in our 
measurement of the $\ttbar$ invariant mass spectrum
must first contain a lepton candidate in the central detector region 
($|\eta|$ $<$ 1.0).  This lepton is required to be either an isolated 
electron with 
transverse energy ($\ET$) in 
excess of 20 ${\rm GeV}$ or an isolated muon with transverse momentum ($\PT$)
in excess of 20 ${\rm GeV/c}$.  
Events must also include at least 20 ${\rm GeV}$ of
$\MET$, attributable to the presence of a neutrino, as well as at least four
jets with $|\eta|$ $<$ 2.0 and raw $\ET$ $>$ 15 ${\rm GeV}$.  Raw jet 
energies are the 
values which result from clustering signals in the 
calorimeter towers 
before any offline jet corrections are applied.
To increase
the acceptance rate for $\ttbar$ events, the requirements for the fourth jet
are relaxed such that the raw $\ET$ must only be greater than 8 ${\rm GeV}$ 
with $|\eta|$ $<$ 2.4 
in events where at least one of the leading three jets is tagged by 
the SVX or SLT algorithms.  All jets in this analysis are formed as clusters of
calorimeter towers within cones of fixed radius $\Delta R \equiv \sqrt{(\Delta
\eta) ^2 + (\Delta \phi) ^2} = 0.4$.  In 106 $\ipb$ of data, we observe 83 
events which satisfy these requirements.


This method builds upon the
techniques developed for the top quark mass measurement\cite{top_prl} by 
fitting each event to the hypothesis of $\ttbar$ production followed by decay 
in the lepton+jets channel:

%
\begin{picture}(120,120)(75,-20)
\put(80.0,80.0){$p \; \bar p \; \longrightarrow$}
\put(137.0,80.0){$t$}
\put(139.0,77.0){\line(0,-1){60.0}}
\put(157.0,80.0){$\bar t$}
\put(177.0,80.0){${\xi}$}
\put(159.0,77.0){\line(0,-1){15.0}}
\put(159.0,62.0){\vector(1,0){30.0}}
\put(192.0,58.0){$W^- \; \; \bar b$}
\put(198.0,55.0){\line(0,-1){15.0}}
\put(198.0,40.0){\vector(1,0){30.0}}
\put(231.0,36.0){$q \; \; {\bar q^\prime}$ (or $l^- \; \; \bar \nu_l$)}
\put(139.0,17.0){\vector(1,0){50.0}}
\put(192.0,13.0){$W^+ \; \; b$}
\put(198.0,10.0){\line(0,-1){15.0}}
\put(198.0,-5.0){\vector(1,0){30.0}}
\put(231.0,-9.0){$l^+ \; \; \nu_l$ (or $q \; \; {\bar q^\prime}$)}

\normalsize
\end{picture}
%
%

\noindent The four-momenta of these 13 objects fully describe a $\ttbar$ event. 
The three-momenta of the charged lepton and four jets are measured directly. 
To compute the energies of these objects, the $b$ and ${\bar b}$ quark masses 
are taken to be 5 $\gev$, the $q$ and ${\bar q^\prime}$ masses are taken to be 
0.5 $\gev$, and the charged lepton mass is 
assigned according to its identification as either an electron or a muon.
The components of transverse momentum for the recoiling system, $\xi $, are 
measured directly from extra jets in the event and 
unclustered energy deposits that are not included in lepton or jet 
energies.  The transverse momentum components of the neutrino are computed by 
requiring that the total $\ET$ in the event sums to zero.  While the neutrino 
is assumed to be massless,
its longitudinal momentum 
is a free parameter 
in the kinematic fit in which the $q{\bar q^\prime}$ and $\ell \nu$ invariant 
masses are constrained to equal the $W$ boson mass.  We perform a kinematic fit
to the production and decay of the $\ttbar$ pair
as described by the decay chain shown above.  This fitting procedure, which
depends on the minimization of a $\chi^2$ expression\cite{prd}, allows the 
lepton energy, the jet energies and the unclustered energy to vary within their
respective uncertainties.  
The fitted results for these values determine the $t$ and ${\bar t}$ 
four-momentum, 
from which the $\ttbar$ invariant mass ($\Mtt$) can be computed.  To improve
the $\Mtt$ resolution, we also constrain the two $Wb$ invariant
masses to 175 GeV/c$^2$, in agreement with the most recent measurement of the 
top quark
mass\cite{cdf_d0}.  We use only 
the four highest $\ET$ jets, leading to 12 combinations for assigning jets to 
the $b$, ${\bar b}$, and hadronic $W$ 
daughters.  However, because we measure only the transverse component of the 
total energy, thereby determining $\MET$, a two-fold ambiguity in the 
longitudinal component of the neutrino momentum results in 24 combinations.
We further require that 
jets that are SVX or SLT-tagged be assigned to $b$-quarks, thereby reducing 
the number of 
combinations.
%
                            
Electron energies and muon momenta 
are measured with the 
calorimeter and tracking chambers, respectively\cite{cal_tr}.  
A set of generic jet corrections is applied to the energies of all the
jets in an event to account for absolute energy scale calibration,
contributions from the underlying event and multiple interactions, as well as
energy losses in cracks between detector components and outside the clustering
cone.
These 
corrections are 
determined from a combination of Monte Carlo simulations and 
data\cite{jetcorr}.  The four leading jets in a $\ttbar$ event undergo an 
additional energy correction that depends on the type of parton that they are 
assumed
to be in the fit: a light quark, a hadronically decaying $b$ quark, or 
a $b$ quark that decayed semileptonically.  These parton-specific corrections
account for (a) the differences in the expected $\PT$ distributions of jets 
from $\ttbar$ and the shape which was assumed to derive the generic jet 
corrections mentioned above, and (b) the energy losses from semileptonic $b$ 
and $c$-hadron decays.  These corrections were derived from a
study of $\ttbar$ events generated with the {\sc herwig} Monte Carlo
program\cite{her}.

Using Monte Carlo simulations of signal and background events, we explored 
several event selection criteria in an attempt to optimize our discovery 
potential\cite{thesis}.  Of the 24 possibilities for each event, we select the 
$\Mtt$ value which corresponds to the 
configuration with the 
lowest $\chi^2$.  
To reduce the probability of selecting
configurations with incorrect parton assignments which tend to yield 
artificially low values of
$\Mtt$, we refit each event after releasing the constraint that
the $Wb$ invariant mass be equal to 175 GeV/c$^2$ and demand that the fit
for this particular configuration return a value for the top quark mass
between 150 GeV/c$^2$ and 200 GeV/c$^2$.
To further reduce incorrect combinations
and to increase discovery potential for a new particle decaying to $\ttbar$,
we apply a $\chi^2$ cut.
For narrow width $\ttbar$ resonances, simulation predicts that the width of
the $\Mtt$ spectrum is $\sim6$\% of the resonance mass for cases in which the
correct jet configuration is selected.  For resonances with a natural width
$\Gamma$ that is significantly less than 6\% of the nominal mass, the CDF
detector resolution will dominate and the resonances will all have
approximately the same shape (shown in the inset of Fig. 1 for a mass of 500 
$\gev$).
At low $\Mtt$, the presence of residual events
with incorrect parton assignments is evident in this figure.

\begin{figure}
\hspace*{0.3cm}
\epsfysize=3.25in
\epsffile[72 162 522 648]{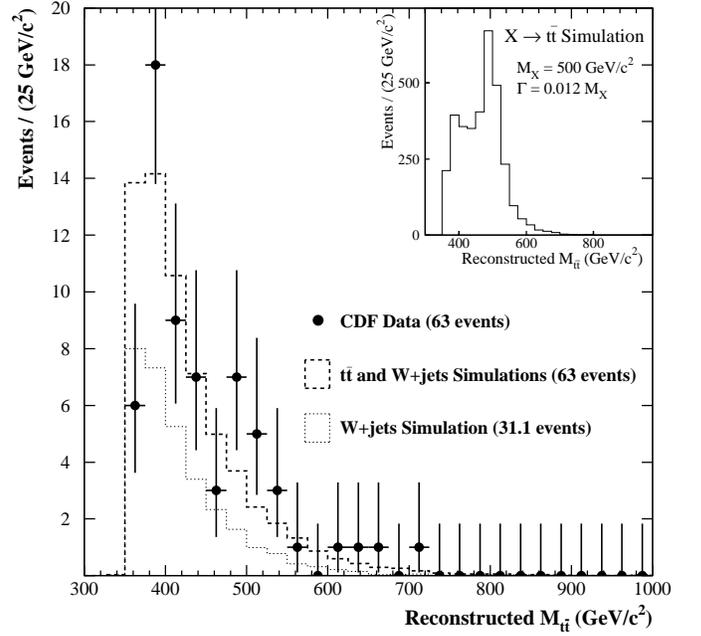}
\vspace{0.25cm}
\caption{The observed $\Mtt$ spectrum (points) compared to the 
QCD $W$+jets background (fine dashes) and the total standard model prediction
including both QCD $W$+jets and $\ttbar$ production (thick dashes).  The 
$\ttbar$ prediction has been normalized such that the number of events in the 
total standard model prediction is equal to the number of events in the 
data.  The inset shows the expected $\Mtt$ shape resulting from the simulation 
of a narrow, vector $X \rightarrow \ttbar$ resonance ($\Mx = 500 \; \gev$, 
$\Gamma = 0.012 \Mx$) in the CDF detector.}
\label{fig:mtt}
\end{figure}

The selection criteria described above eliminate an additional 20 events from 
our data
sample and the resulting $\Mtt$ spectrum is shown in Fig. \ref{fig:mtt}, along
with the expected standard model $\ttbar$ and QCD $W$+jets background shapes
normalized to the data.  While the non-$\ttbar$ background is dominated by 
$W$+jets events, it also includes contributions from multijet $\bbar$ 
events with one jet misidentified as a lepton, $Z$+jets events, events with a 
boson pair, and single-top production.  However, it has been 
shown that the {\sc vecbos}\cite{vec} $W$+jets shape alone is sufficient in 
modeling the entire non-$\ttbar$ background spectrum\cite{top_prl}.  For this 
analysis, the 
expected non-$\ttbar$ background prediction of 31.1 $\pm$ 8.5 events is 
calculated as in
Ref. \cite{prd}, 
but accounts for differences in selection criteria.  
We find that the $\Mtt$ distribution of 63 data events
is consistent with
the hypothesis that the spectrum is comprised of 
standard model $\ttbar$ production and the predicted rate of non-$\ttbar$ 
background events, as shown in Fig. \ref{fig:mtt}.                            

Because we cannot present evidence for a narrow $\ttbar$ resonance, we
establish upper limits on the production cross-section for a new vector 
particle, 
$X$, of mass $\Mx$ decaying and to 
$\ttbar$.  For natural widths $\Gamma = 0.012\Mx$ and $\Gamma = 0.04\Mx$, and 
for 
each $\Mx$ between 400 $\gev$ and 1 $\tev$ in increments of 50
$\gev$, we perform a binned-likelihood fit of the data.  To determine the
likelihood function for a given $\Mx$ and $\Gamma$, 
we fit the $\Mtt$ spectrum from the data to the 
expected Monte Carlo shapes for both the $\ttbar$ and QCD $W$+jets background
sources as well as a resonance signal $X \rightarrow \ttbar$ which 
we model using $\zpr \rightarrow \ttbar$ in {\sc pythia}\cite{pyt}.  
Our analysis is subject to several sources of systematic
uncertainty on the
expected shape of background and signal $\Mtt$ spectra and/or the signal
acceptance rate.
Treating these two types of 
systematic effect separately, we establish the magnitude of each source 
through a Monte Carlo procedure which quantifies the effect of varying the 
source of uncertainty by one standard deviation.  
We determine the uncertainty contributions due to the jet $\ET$ scale,
initial and final state gluon radiation, and the non-$\ttbar$ background
spectrum using methods described in Ref. \cite{prd}.  
The uncertainty in the measurements of the 
top quark mass\cite{cdf_d0} and total integrated luminosity\cite{lum} 
are included in our study of systematic effects, as well as the uncertainty due 
to the choice of parton distribution functions (PDF).  The remaining sources of
systematic uncertainty considered are all small and include trigger 
efficiency, lepton 
identification, tracking efficiency, $z$-vertex efficiency, and Monte Carlo 
statistics.  
The uncertainties resulting from jet $\ET$ scale
and top quark mass are correlated and we conservatively take this correlation
to be 100\%.

The percent uncertainty in 
$\sigma_{\tiny {\mbox X}} \cdot \mbox{BR}\{{\mbox X} \rightarrow \ttbar\}$
is listed in Table \ref{table:sys} for each of the systematic sources at 
several different resonance masses.
The systematic effect due to uncertainty in the top quark mass ($M_{top}$) is 
dominant at low $\Mx$, whereas the effect due to the 
uncertainty in modeling final state radiation dominates at large 
$\Mx$.  
To ensure that our estimates are conservative, the systematic uncertainty
is taken to be a constant number of pb below the value of $\sbr$ 
corresponding to the 95\% C.L. limit obtained with statistical uncertainties 
only\cite{thesis}.  That
constant is the estimate of the systematic uncertainty at the 95\% C.L.
limit.  Above the same value of $\sbr$, we use a systematic uncertainty that
rises with $\sbr$ at the fixed percent rate listed in Table \ref{table:sys}.
        
\begin{figure}
\hspace*{0.3cm}
\epsfysize=3.25in
\epsffile[72 162 522 648]{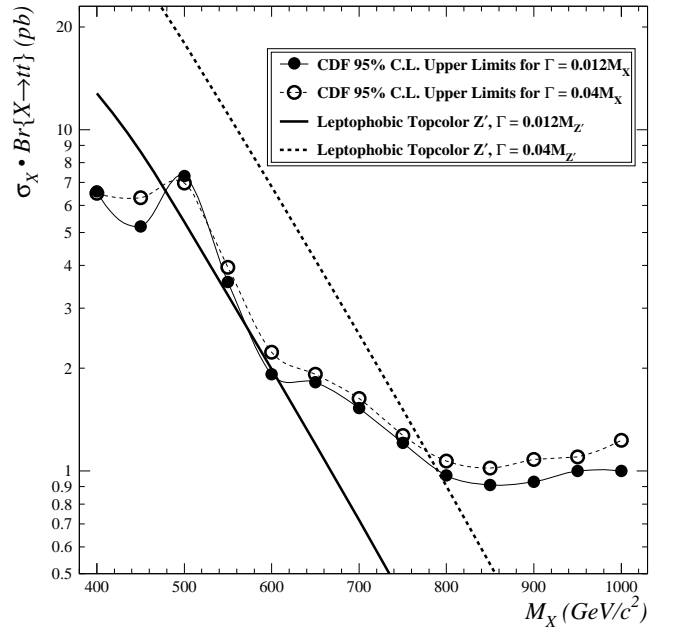}
\vspace{0.25cm}
\caption{
The 95\% C.L. upper limits on $\sbr$ as a function of mass
(solid and open points) compared to the cross section for a leptophobic
topcolor $\zpr$ (thick solid and dashed curves) for two resonance widths
($\Gamma$ = 0.012 $\Mz$ and $\Gamma$ = 0.04 $\Mz$).}
\label{fig:lim}
\end{figure}

For each resonance mass and width, we convolute the statistical likelihood 
shape with the Gaussian total systematic 
uncertainty and extract the 95\% C.L. upper limit on 
$\sigma_{\tiny {\mbox X}} \cdot \mbox{BR}\{{\mbox X} \rightarrow \ttbar\}$
which is listed in Table \ref{table:results} and shown in Fig. \ref{fig:lim}.
The systematic uncertainties increase the 95\% C.L. upper limit by 27\% for
$\Mx$ = 400 $\gev$, but only 7\% (6\%) for $\Mx$ = 600 (800) $\gev$ because
statistical uncertainties dominate the likelihood.
Also shown in Fig. 
\ref{fig:lim} are the theoretical 
predictions for cross-section 
times branching ratio for a 
leptophobic $\zpr$
with natural width 
$\Gamma$ = $0.012\Mz$ and $\Gamma$ = $0.04\Mz$\cite{models}.  
At 95\% confidence, we exclude the existence of a leptophobic topcolor $\zpr$ 
with mass $\Mz$ $<$ 480 $\gev$ for natural width $\Gamma$ = 0.012 $\Mz$, and  
mass $\Mz$ $<$ 780 $\gev$ for $\Gamma$ = 0.04 $\Mz$.  
For larger widths, detector resolution will no longer be the
dominant factor in determining the $\zpr$ signal shape, so our limits are no
longer applicable.  

In conclusion, after investigating 106 $\ipb$ of data 
collected at CDF, we find no evidence for a $\ttbar$ resonance and establish
upper limits on cross-section times branching ratio for narrow resonances.  
We have used these limits to
constrain a model of topcolor assisted technicolor.
We thank the Fermilab staff and the technical staffs of the participating
institutions for their vital contributions.  This work is supported 
by the U.S. Department of Energy
and the National Science Foundation; the Natural Sciences and Engineering
Research Council of Canada; the Istituto Nazionale di Fisica Nucleare of
Italy; the Ministry of Education, Science and Culture of Japan; the National
Science Council of the Republic of China; and the A.P. Sloan Foundation.

%
%
\begin{table}
\begin{center}
\begin{tabular}{c c c c c}
\put(0,10){\line(5,-1){70.0}} & $\Mx$ (GeV/c$^2$) & 400 & 600 & 800 \\ 
Source & \put(27,10){\line(5,-1){70.0}} & & & \\ \hline
\multicolumn{2}{l}{Jet $\ET$} &   6.1    &   6.2    &   4.4 \\
\multicolumn{2}{l}{$M_{top}$} &   22     &   3.1    &   8.7 \\
\hline
\multicolumn{2}{l}{Jet $\ET$ and $M_{top}$} &   28     &   9.3    &   13  \\
\hline
\hline
\multicolumn{2}{l}{Initial state radiation} &   14     &   4.2    &   5.6 \\
\multicolumn{2}{l}{Final state radiation} &   19     &   16     &   12  \\
\multicolumn{2}{l}{$b$-tagging bias} &   4.6    &   0.79   &   0.85  \\
\multicolumn{2}{l}{PDF} &   11     &   5.5    &   4.8 \\
\multicolumn{2}{l}{QCD background shape} &   1.3    &   0.17   &   0.045 \\
\multicolumn{2}{l}{Additional acceptance effects} &  5.3 & 5.3 &   5.3 \\
\multicolumn{2}{l}{Luminosity} &   4.0    &   4.0    &   4.0 \\
\hline
\hline
Total
        & & 39 & 21 & 20 \\
\end{tabular}
\end{center}
\caption{The percent systematic uncertainty in $\sbr$ from various sources.}
\label{table:sys}
\end{table}

\begin{table}
\begin{center}
\begin{tabular}{c cc } 
\multicolumn{1}{c}{$\Mx$}&  
\multicolumn{2}{c}{95\% C.L. upper limit} \\ 
\multicolumn{1}{c}{$(\gev)$} &  
\multicolumn{2}{c}{$\sbr$ (pb)} \\ 
\hline
   & for $\Gamma = 0.012 \Mx$ & for $\Gamma = 0.04 \Mx$ \\
\hline
400 & 6.60 & 6.51 \\
450 & 5.21 & 6.32 \\
500 & 7.31 & 6.97 \\
550 & 3.58 & 3.95 \\
600 & 1.92 & 2.23 \\
650 & 1.82 & 1.92 \\
700 & 1.53 & 1.63 \\
750 & 1.21 & 1.27 \\
800 & 0.97 & 1.07 \\
850 & 0.91 & 1.02 \\
900 & 0.93 & 1.08 \\
950 & 1.00 & 1.10 \\
1000 & 1.00 & 1.23 \\
\end{tabular}
\end{center}
\caption{The 95\% C.L. upper limit on the cross section times branching ratio
for
vector particles decaying to $\ttbar$, as a function of mass, for two natural
widths.}
\label{table:results}
\end{table}


\begin{references}
\bibitem{met_def} The transverse momentum of a particle is $\PT = P sin\theta$.
The analogous quantity using calorimeter energies, defined as
$\ET = E sin\theta$, is called transverse energy.
Missing transverse energy $\MET$ is defined as
$-\sum \ET^i \,\hat{n_i}$, where $\hat{n_i}$ are the
unit vectors pointing from the interaction point to
the energy deposition in the calorimeter.
\end{references}
\end{document}